\documentstyle[pra,aps,multicol,graphicx]{revtex}

\begin{document}

\draft

\title{Symmetry Violation in a Superconducting Film \\
with a Square Array of Ferromagnetic Dots.}

\author{Serkan Erdin}

\address{School of Physics \& Astronomy, University of Minnesota,\\
Minneapolis, MN 55455}

\date{\today}
\maketitle
\begin{abstract}
We study the equilibrium state of a superconducting film covered
with a regular square array of perpendicularly magnetized magnetic
dots. The dots induce vortices in the film directly under them and
antivortices between the dots. We show that the symmetry of the
dot array is spontaneously violated by the vortices. The positions
of the vortices and the antivortices depend on the magnetization
and the size of the dots.
\end{abstract}

\begin{multicols}{2}
%\narrowtext
Heterogeneous structures made of ferromagnetic (FM) and
superconducting (SC) pieces have attracted a great deal of
interest both experimentally \cite{e1,e2,e4,e5,e6,e7,e8} and
theoretically in recent years 
\cite{pok1,pl1,pok3,feldman,helselth,mar,san1,amin}. In
contrast to conventional systems, these structures allow the
coexistence of ferromagnetism and superconductivity, if FM and SC
order parameters are separated in space by a thin insulating layer
\cite{pok1}. Strong interaction of the FM and SC systems not only
gives rise to a new class of novel phenomena and physical effects,
but also shows the important technological promise of devices
whose transport properties can be easily tuned by comparatively
weak magnetic fields.

Several realizations of such systems were proposed. However, so
far only sub-micron magnetic dots covered by thin SC films have
been prepared and studied \cite{e1,e2,e4,e5,e6,e7,e8}. The
experimental samples of FM-SC hybrid systems were prepared by
means of electron beam lithography and lift-off techniques
\cite{ee}. Both in-plane and out-of-plane magnetization was
experimentally realized. The dots with magnetization parallel to
the plane were fabricated from Co, Ni, Fe, Gd-Co and Sm-Co alloys.
For the dots with magnetization perpendicular to the plane,
experimenters used Co/Pt multilayers. The FM dots were deposited
on thin SC films made of either Nb or Pb, whose transition
temperatures are around $7-10$ K. In these experiments, the
effects of commensurability on the transport properties, e.g. the
magnetoresistance oscillations and matching anomalies were
observed. Further experiments with periodic arrays of magnetic
dots or holes \cite{met1,met2,met3,met4}  confirm that the FM dots
create and pin vortices.

Motivated by current interest and appeal of dealing with a new
class of physical systems, we consider a periodic array of FM dots
on SC thin film. We assume that not only the location of dots is
regular, but also their magnetic moments are parallel. Such
magnetically ordered dot array can be obtained by the
field-cooling from magnetic Curie temperature to a temperature
below the superconducting transition temperature $T_s$. It is
worth to mention that zero-field cooling leads to either random
dot magnetization for magnetically hard dots, or to an
antiferromagnetic ordering if the dots are magnetically soft
\cite{pok1,feldman}.
%Here you should write that, as we demonstrated in common work, a
%single dot with sufficiently strong magnetization and large radius
%generates a vortex or several vortices under itself. In a periodic
%array of dots they must be compensated by an array of antivortices,
%since the total flux generated by a vortex is zero. Discuss where
%the antivortices can be located. Thus, the main problem is to find
%the number and locations of vortices and antivortices in the
%elementary cell. Discuss why symmetric positions may be energy
%unfavorable. Say what will be demonstrated.

In our previous work \cite{common}, we demonstrated that a single FM dot 
with  sufficiently strong magnetization and large radius
generates a vortex or several vortices under itself. In a periodic
array of dots they must be compensated by an array of antivortices,
since the total flux generated by a vortex is zero. Depending on the dot's 
magnetization and size, there might exist several vortex-antivortex pairs 
in the 
array's elementary cell. For simplicity, we consider only  one pair. Due 
to the symmetry of a square array, the vortices are expected to sit at the 
dot's centers, while the antivortices are located at the centers of the 
elementary cells. However, symmetric positions may be energetically 
unfavorable due to the inhomogenaous magnetic field distribution over the 
film. Thus, the main problem is to find positions of a vortex and an 
antivortex in an elementary cell.

In this system, the magnetic field induced by inhomogeneous
magnetization of the dot array penetrates into a superconductor
and generates the SC vortices, whereas the magnetic field
generated by the supercurrents and SC vortices acts on the
magnetic system. The method based on London-Maxwell equations to
study such a system was developed in \cite{common}. Under the
assumption that the sizes of all structures in the problem are
larger than the coherence length $\xi$, the London approximation
is valid.  In section I, we apply the method of the work
\cite{common} to periodic FM-SC structures. Next, we apply it to a
system in which a square array of thin circular FM dots is placed
upon a thin SC film. We conclude with the results and discussion.

\section{Energy and magnetic field in periodic heterogeneous FM-SC systems}

A periodic heterogeneous FM-SC system such as the regular magnetic
dot array upon the SC film or a periodic domain structure in a
ferromagnet-superconductor bilayer can be studied with the method
described in \cite{common}.The method to study the
ferromagnet-superconductor bilayer in the continuous limit, where the 
magnetic system's size is larger than effective penetration depth, is 
given in  
\cite{bilayer}. In this section, we introduce a more general method which 
works well both in continuous and discreet limits.   
%We did it in our work about bilayer. What new is made in this section?
%Please, mention this our calculation and be more specific.
%In this section, we extend the aforementioned method to study
%periodic structures by means of Fourier series. 
In doing so, we assume 
that the periodic
structures of interest are made of very thin magnetic textures
with the magnetization perpendicular to the plane and SC films. For 
simplicity, we consider the limit of zero thickness for films with 
non-zero 2d magnetization
and density of superconducting electrons.  
Their energy is calculated 
over the surface of 
the SC 
film. 
%I do not understand what do you mean. Does it mean that you will
%consider the limit of zero thickness, but non-zero 2d magnetization
%and density of superconducting electrons?
We start with the energy of such 2d systems:
\begin{equation}
U = U_{vv} + U_{mv} + U_{mm},
\label{en1}
\end{equation}
where $U_{vv}$ is the vortex energy, $U_{mv}$ is the energy of
magnetization-vortex interaction, and $U_{mm}$ is the magnetic
energy.  The terms in (\ref{en1}) are as follows (see
\cite{common} for details):
\begin{eqnarray}
U_{vv} &=& \int [\frac{\varepsilon_0}{2 \pi} ({\bf \nabla} \varphi )^2 -
\frac{\phi_0}{16 \pi^2 \lambda} ({\bf \nabla} \varphi ) \cdot {\bf a}^{v}
] d^2 r, \label{en-v} \\
U_{mv} &=& - \int [\frac{\phi_0}{16  \pi^2 \lambda}  ({\bf \nabla} \varphi
\cdot {\bf a}^{m} - \frac{1}{2} {\bf b}^{v} \cdot {\bf m} ] d^2 r,
\label{en_mv} \\
U_{mm} &=& - \frac{1}{2} \int {\bf b}^{m} \cdot {\bf m} d^2 r
\label{en_mm},
\end{eqnarray}
where $\varepsilon_0 = \phi_0^2/16 \pi^2 \lambda$, $\phi_0 = h c/
2 e$ is the quantum flux, $\lambda = \lambda_L^2/d_s$ is the
effective penetration depth \cite{abrikosov},  $({\bf \nabla} \varphi)$ is 
the
phase gradient induced by the vortex lattice, ${\bf a}$ and ${\bf
b}$ are vector potential and magnetic field at $z=0$,
respectively, ${\bf m}$ is the 2d magnetization. The superscripts
$m$ and $v$ indicate the contributions of magnetization and
vortices, respectively. In periodic system all these values are
periodic functions of coordinates and can be expanded into
Fourier-series. For any periodic 2d function ${\bf f}({\bf r})$,
the Fourier expansion is given by
\begin{equation}
{\bf f} ({\bf r}) = \sum_{\bf G} {\bf f}_{\bf G} e^{{\bf G} \cdot {\bf r}}
\hspace{1cm} {\bf f}_{\bf G} = \frac{1}{\cal A} \int {\bf f} ({\bf r})
e^{{\bf G} \cdot {\bf r}} d^2{\bf r}.
\label{fourier}
\end{equation}
where ${\bf G}$ are the reciprocal vectors of the periodic system
and ${\cal A}$ is the unit cell area. Representing all the values
${\bf a}$, ${\bf b}$, ${\bf m}$ and ${\bf \nabla} \varphi$ as
Fourier-series (\ref{fourier}), plugging them into the integral
representation of the energies
(\ref{en-v},\ref{en_mv},\ref{en_mm}) and using equality $\int
e^{i({\bf G} + {\bf G}^\prime)\cdot{\bf r}} d^2 r = {\cal A}
\delta_{{\bf G},-{\bf G}^\prime}$, we obtain the following
expression for the energy per unit cell:

\begin{eqnarray}
u_{vv} &=& \sum_{\bf G}\left[\frac{\varepsilon_0}{2 \pi} |({\bf
\nabla} \varphi)_{\bf G}|^2 - \frac{\phi_0}{16 \pi^2 \lambda} {\bf
\nabla}
\varphi_{\bf G} \cdot {\bf a}_{-\bf G}^{v}\right], \label{puv} \\
u_{mv} &=& - \sum_{\bf G} \left[\frac{\phi_0}{16 \pi^2 \lambda}
{\bf \nabla} \varphi_{\bf G} \cdot {\bf a}_{-\bf G}^{m} +
\frac{1}{2} {\bf b}_{\bf
G}^{v} \cdot {\bf m}_{-{\bf G}}\right], \label{pumv} \\
u_{mm} &=& - \frac{1}{2} \sum_{\bf G} {\bf b}_{\bf G}^{m} \cdot
{\bf m}_{-{\bf G}}. \label{pumm}
\end{eqnarray}

The total vector-potential obeys the London-Pearl equation
\begin{equation}
\nabla^2 {\bf A} = \frac{1}{\lambda} {\bf a} \delta (z) -
4 \pi {\bf \nabla} \times ({\bf m}\delta(z)) - \frac{\phi_0}{2 \pi
\lambda} ({\bf \nabla} \varphi ) \delta ( z ). \label{lonpearl}
\end{equation}
To find the vector-potentials generated by magnetization and
vortices, we employ the Fourier-expansions of ${\bf a}$, \\
$({\bf\nabla} \varphi)$ and ${\bf m}$. In this article, we are interested 
only in the FM subsystems with magnetization in the $z$ direction. The 
calculations are therefore done along with the Fourier coefficient of 
$m_z$. The Fourier coefficients of the
phase gradient are given by (see, for example \cite{abrikosov})
\begin{equation}
({\bf \nabla} \varphi)_{\bf G} = \frac{2 \pi}{\cal A}
\frac{i {\bf G} \times \hat z F_{\bf G}}{G^2}, \label{phase}
\end{equation}
where $F_{\bf G} = \sum_i n_i e^{i {\bf G} \cdot {\bf r}_i}$ is
the structure factor of vortices, $n_i$ and ${\bf r}_i$ indicate
the vorticity and the position of the $i$-th vortex respectively.
In doing so, we get the Fourier coefficients of the vector-
potentials generated by magnetization in the $z$ direction and vortices:
\begin{eqnarray}
{\bf a}_{\bf G}^{v} &=& \frac{\phi_0}{\cal A} \frac{i ({\bf
G} \times \hat z)F_{\bf G}}{ G^2 ( 1 + 2 \lambda G )},\label{av} 
\\
{\bf a}_{\bf G}^{m} &=& 4 \pi \lambda \frac{i ({\bf G} \times \hat z) m_{z
{\bf G}}}{1 + 2 \lambda G}.\label{am}
\end{eqnarray}
Fourier-image of the magnetic field at the interface is ${\bf
b}_{\bf G} = i {\bf G} \times {\bf a}_{\bf G}$. Substituting the
Fourier coefficients of the vector-potentials, magnetic fields,
the phase gradient and the magnetization into equations
(\ref{puv},\ref{pumv},\ref{pumm}), we find the energy of the
periodic structure:
\begin{eqnarray}
u_{vv} &=& \frac{\phi_0^2}{4 \pi{\cal A}^2} \sum_{\bf G}
\frac{ |F_{\bf G}|^2}{G ( 1 + 2 \lambda G)}, \label{hv} \\
u_{mv} &=& -
\frac{\phi_0}{\cal A} \sum_{\bf G} \frac{m_{z {\bf G}}F_{-\bf G}}{1 + 2
\lambda G}, \label{hmv}\\
u_{mm} &=& -2 \pi \lambda \sum_{\bf G} \frac{G^2
|{\bf m}_{z\bf G}|^2}{1 + 2 \lambda G}. \label{hmm}
\end{eqnarray}
These equations have rather general character. The only
assumptions are that magnetic and superconducting subsystems are
2-dimensional and periodic.

\section{The Square Array of FM Dots}

\begin{figure}[h]
\begin{center}
\includegraphics[angle=0,width=3.5in,totalheight=3.5in]{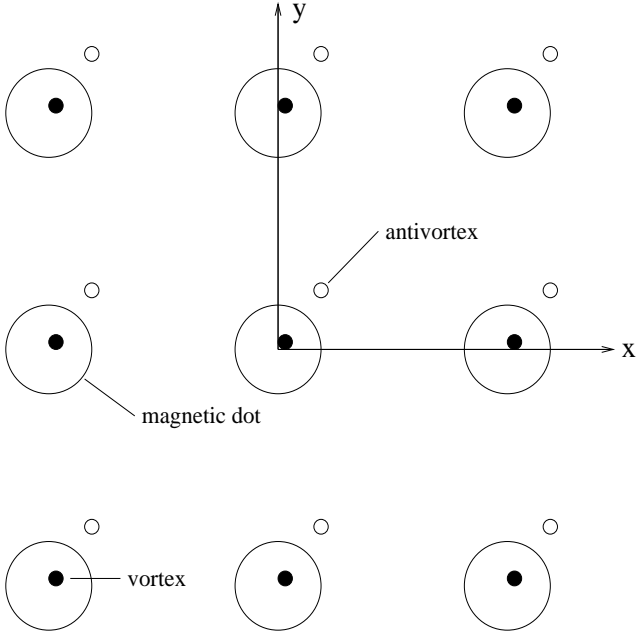}
\caption{\label{mda} The top view of the square magnetic dot
array is as above. The vortices are confined  within  the dot's region,
while the antivortices appear outside.}
\end{center}
\end{figure}

In this section we consider a square array of the FM dots on a SC
film. We assume that criterion of the vortex appearance under a
single magnetic dot \cite{common} is satisfied. Then the vortices
appear under the FM dots, while antivortices appear outside them.
The existence of antivortices is obvious since the energy of a
system containing only vortices would grow as the cube of the
system's linear size, whereas, in the "neutral" vortex-antivortex
system the energy grows as the square of its size. We consider a
simple case in which each dot creates only one vortex under them
and one antivortex outside (see Fig.\ref{mda}). In this section we
find the condition for the appearance of the vortex-antivortex
state and their locations minimizing the energy. In order to solve
these problems, we follow the theoretical procedure outlined in
the previous section.

The geometry of the FM dots is assumed to be the same as that in
the work \cite{common}, namely the magnetization of each dot
points in positive $z$-direction, they are circular with the
radius of each dot equal to $R$.  The lattice constant of the
square array is denoted as $L$. The SC film and the FM dots are
placed at the heights $z=0$ and $z=d>0$, respectively. Later, we
transfer to the limit $d=0$ to simplify the results.
For  a square lattice with the unit cell containing a
vortex-antivortex pair, the form factor is:

\begin{equation}
F_{\bf G} = \exp{(i {\bf
G} \cdot {\bf r}_1)} -  \exp{(i {\bf G} \cdot {\bf r}_2)}.
\label{sf}
\end{equation}
\noindent The set of the reciprocal lattice vectors for the square
array is given by equation: ${\bf G} = ( 2 \pi n/L, 2 \pi s/L)$,
where $n,s$ are integers; ${\bf r}_1 = x_1 \hat x + y_1 \hat y$
and ${\bf r}_2 =  x_2 \hat x + y_2 \hat y$ are the position
vectors of the vortex and the antivortex, respectively. In order
to find the energy of the system, we need the Fourier-coefficients
of the magnetization. Employing the definition of Fourier
coefficients for periodic systems in (\ref{fourier}), the
magnetization of a square array of FM dots reads
\begin{equation}
m_{z {\bf G}}= \frac{2 \pi m R}{L^2} \frac{J_1 ( G R )}{G}.
\label{fcmag}
\end{equation}
In this result the distance between the dots and SC film $d$ is
taken to be zero.

We compute  the vortex energy first. Substituting
(\ref{sf}) into Eq.(\ref{hv}) gives the energy per unit cell:

\begin{equation}
u_{vv} = \frac{8 \pi  \varepsilon_0 \lambda}{L^4} \sum_{\bf G}
\left[\frac{1}{G
( 1 + 2
\lambda G )} - \frac{\cos({\bf
G}
\cdot ({\bf r}_2 - {\bf r}_1 ))}{G ( 1 + 2 \lambda G )}\right].
\label{perH_v}
\end{equation}
\noindent The cosine  term in (\ref{perH_v}) gives the vortex
interaction energy; whereas, the first term is the vortex self
energy. In order to extract the logarithmic contribution of the
vortex self energy,  we split the first term in two parts as $(4
\pi\varepsilon_0/L^4) \sum_{\bf G} [1/G^2 - 1/(G^2 (1+ 2 \lambda
G))]$. The first term is logarithmically divergent and equals $(2
\varepsilon_0/L^2) \ln (L/\xi)$. The second term is convergent and
will be left in  series form. Thus, the  vortex energy is found
as:
\begin{equation}
u_{vv} = \frac{2 \varepsilon_0}{L^2} \bigl[\ln \frac{\lambda}{\xi} -
\frac{1}{2\pi} f_v(\lambda /L)
- \frac{2 \lambda}{L} f_{vv} (\lambda/L,{\bf
r}_1/L,{\bf r}_2/L )\bigl].
\label{hvper}
\end{equation}

\noindent where the functions $f_v(\lambda/L)$
$f_{vv}(\lambda/L,{\bf r}_1/L,{\bf r}_2/L)$ are defined by series:

\begin{eqnarray}
f_v &=& {\sum_{n,s =-\infty}^{\infty \; \prime}}  \frac{1}{(n^2 + s^2)(1
+ 4 \pi
\frac{\lambda}{L} \sqrt{n^2 + s^2})}, \nonumber \\
f_{vv} &=&{\sum_{n,s =-\infty}^{\infty \; \prime}}  \frac{\cos ( 2
\pi (n (x_2 - x_1)/L + s (y_2 - y_1)/L))}{\sqrt{n^2 + s^2}(1 + 4 \pi 
\frac{\lambda}{L}\sqrt{n^2 + s^2})},
\label{fvvp}
\end{eqnarray}
where the notation $\sum^\prime$ indicates that the term with
$n=s=0$ must be omitted.

The magnetization-vortex interaction energy can be obtained by
inserting (\ref{fcmag}) into equation (\ref{hmv}). The result is:
\begin{equation}
u_{mv} = -\frac{ m \phi_0 R}{L^3} f_{mv} ( \lambda/L,R/L,{\bf
r}_1/L,{\bf r}_2/L ), \label{arrayhmv}
\end{equation}.
\end{multicols}
\noindent where,
\widetext{
\begin{equation}
f_{mv} = {\sum_{n,s=-\infty}^{\infty \; \prime}} \frac{J_1 ( 2 \pi
\frac{R}{L} \sqrt{n^2+s^2}) \left(
\cos ( 2 \pi (x_1 n + y_1 s )/L ) - \cos ( 2 \pi (x_2 n + y_2
s )/L )\right)}{\sqrt{n^2+s^2}(
1 + 4 \pi \frac{\lambda}{L} \sqrt{n^2 + s^2} )}.
\label{perfmv}
\end{equation}}
\begin{multicols}{2}
Now we are in position to calculate the equilibrium locations of
the vortices and the antivortices. For this purpose the total
energy of the system should be minimized with respect to ${\bf r}_1$ and 
${\bf r}_2$. For this minimization the
magnetization-magnetization interaction can be omitted since it
does not depend on the vortex coordinates. Therefore, only the sum

\begin{equation}
u = u_{vv} + u_{mv}
\label{enper}
\end{equation}
must be minimized.

\begin{figure}[t]
\begin{center}
\includegraphics[angle=0,width=3.5in,totalheight=3.1in]{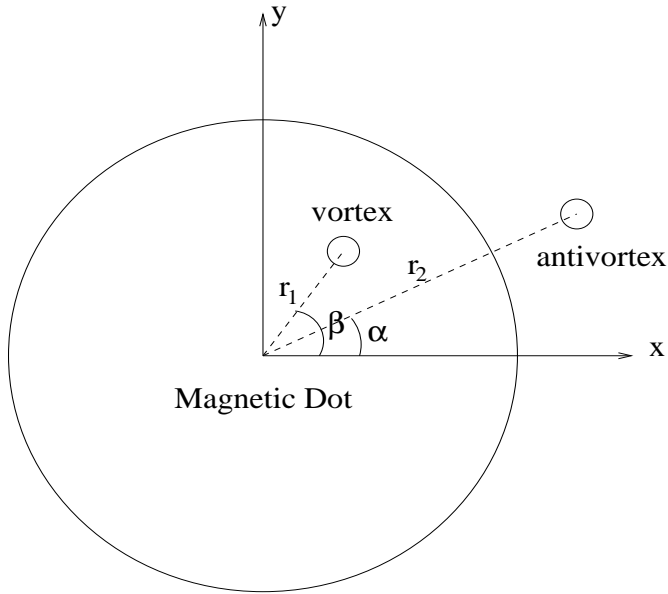}
\caption{\label{coord} The test dot with the vortex and the antivortex.}
\end{center}
\end{figure}
\noindent For numerical analysis, it is convenient to rewrite the
energy in terms of dimensionless parameters. To this end, we
introduce dimensionless variables, namely $\tilde \lambda =
\lambda /L$ , $\tilde R = R /L$, $\tilde {\bf r}_1 = {\bf r}_1/L$
and $\tilde {\bf r}_2 = {\bf r}_2/L$. We also renormalize the
total energy per cell $u$ by dividing it by  $\varepsilon_v/L^2$.
We place  the center of the test dot at the origin, and the
positions of the vortices and the antivortices are calculated with
respect to that point. Let $(\tilde x_1, \tilde y_1)$ and $(\tilde
x_2, \tilde y_2)$ be the respective rescaled  coordinates  of the
vortex and the antivortex relative to the center of the FM dot. 
%We also assume that the vortex and the antivortex are located on a
%straight line passing though the center of the FM dot (see
%Fig.\ref{coord}). This approximation works well as long as the antivortex 
%does not deviate from the line by angles more than $10^0$. 
%Within this range, the change in energy is less than %$10$ of the 
%initial energy. 

%You should discuss this assumption. If it is made only for
%simplicity purpose, you should admit that. Try to deviate ,
%to say, antivortex along the azimuth very little by $\delta|alpha$
%and calculate the linear change of energy. If the coefficient at
%$\delta\alpha$ in energy shift divided by the energy is small, then your
%approximation is good. Otherwise we should think a little more about this.
With these modifications, we minimize (\ref{enper}) numerically
with respect to $\tilde {\bf r}_1$ and  $\tilde {\bf r}_2$ for
different values of $m \phi_0/\varepsilon_v$,  $\tilde R$ and
$\tilde \lambda$. In our numerical calculations, we take
$\lambda/\xi = 50$. In order for the vortices and the antivortices
to appear, the effective energy (see Eq.(\ref{enper})) must become
negative. Using this condition, we find the phase diagrams for the
vortex states, which depend on $R/L$, $R/\lambda$ and  $m
\phi_0/\varepsilon_v$. The ratio

\begin{figure}[t]
\begin{center}
\includegraphics[angle=270,width=3.5in,totalheight=3.5in]{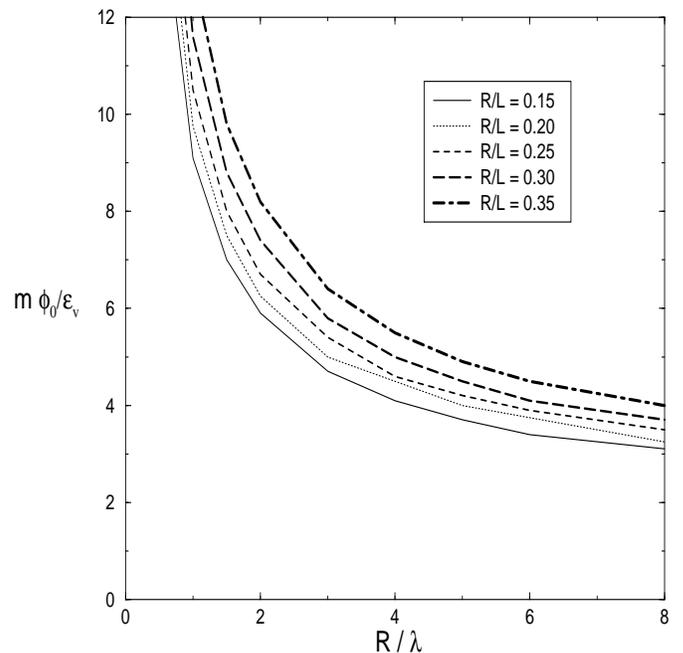}
\caption{\label{1v1a} The phase diagrams of the vortex and the antivortex
state for different $R/L$.}
\end{center}
\end{figure}

\vskip30mm

\begin{equation}
\frac{m \phi_0}{\varepsilon_v} = S g \frac{2 n_m d_m}{n_s ( T ) d_s
\ln \kappa (T)}
\label{coupling}
\end{equation}
\noindent is the relative strength of the SC/FM and controlled by
temperature.
In
Eq.(\ref{coupling}), $n_m$ is the density of magnetic atoms, $d_{m}$ and
$d_s$ are
the respective thicknesses of magnetic and SC films. $S$ is the value of
an elementary spin in the magnet and  $g$ is the Lande factor and equal to
$2$. $n_s (T) = n_{s}(T=0) (1 - T^2/T_s^2)$ is the density of
super-carriers and $\kappa (T) = \kappa(T=0)/\sqrt{1-T^2/T_s^2}$, where
$\kappa(T=0) = \lambda_L^2(0)/d_s \xi (T=0)$. $\ln \kappa(T)$ varies in
the range $3-6$. For typical values of $n_s (0) = n_m = 10^{22} cm^{-3}$,
$d_m = 2 d_s$ , $S = 2$ and near the SC transition temperature $T/T_s
\sim 0.95$, the ratio in (\ref{coupling}) can be around $30$.
The curves in Fig.\ref{1v1a}
indicate when a vortex and an
antivortex appear for different  $R/L$ ratios. Each curve separates the
regions with and without vortex-antivortex pairs. Namely, the region below
the curve for  particular $R/L$ value represents the region without
vortex-and antivortex pair. Above the curve, there exists
vortex-antivortex pairs.
As seen in the figure, the larger the $R/L$ value, the bigger
the $m\phi_0/\varepsilon_v$ ratio. In addition, to create
vortex-antivortex pair spontaneously, the minimum value of
$m\phi_0/\varepsilon_v$ increases as $R/\lambda$ ratio decreases for each
particular ratio  of $R/L$.
Our simulation shows that when the vortices and
antivortices appear on the lines which pass through the centers of the
dots
and make angles $\beta$ and  $\alpha$ with the horizontal axis 
respectively. $\beta$ varies between $3\pi/4$ and $\pi$, while 
$\alpha$ varies between $\pi/4$ and
$\pi/3$, depending on the size of the dot, the period of the array and the
magnetization. It turns out that $\alpha$ gets closer to $\pi/3$ as $R/L$
increases, whereas $\beta$ approaches $\pi$. As the dots are placed 
further away from one another, the vortex moves closer to the antivortex 
and the dot's center.  
 Some
results are
given in the Table \ref{posit}.

\end{multicols}

\begin{table}[h]
\caption{The position of vortices and the antivortices for different
values of the $R/\lambda$ and $m\phi_0/\varepsilon_v$. The three columns
on the left are input. \label{posit}}
\begin{center}
\begin{tabular}{|c|c|c|c|c|c|c|}\hline
{$R/L$}&{$R/\lambda$}&{$m\phi_0/\varepsilon_v$}&{$\tilde r_1$}&{$\tilde 
r_2$}&{$\beta$}&{$\alpha$} \\ \hline
0.35 & 4 & 9  & 0.29  & 0.41 & $180^0$ & $58^0$   \\ \hline
0.30 & 4 & 9  & 0.24  &  0.36 & $174^0$ & $56^0$   \\ \hline
0.25 & 4 & 9  & 0.20 &  0.31 &$166^0$ & $53^0$ \\ \hline
0.20 & 4  & 9 & 0.14 & 0.25 & $161^0$ & $51^0$  \\ \hline
0.15 & 4 & 9  & 0.06 &  0.19 & $157^0$ & $49^0$ \\ \hline
0.35 & 2 & 9 & 0.29 &  0.41  & $178^0$ & $58^0$  \\ \hline
0.30 & 2  & 9 & 0.24  & 0.36 & $171^0$ & $56^0$  \\ \hline
0.25 & 2 & 9 & 0.19 & 0.31 & $164^0$ & $53^0$ \\ \hline
0.20 & 2 & 9 & 0.12 & 0.25 & $159^0$ & $52^0$ \\ \hline
0.15 & 2 & 9 & 0.04 &  0.20 &  $148^0$ & $48^0$  \\ \hline
0.35 & 1 & 15 & 0.28 &  0.41  & $178^0$ &  $58^0$  \\ \hline
0.30 & 1 & 15 & 0.23 & 0.36 & $171^0$ & $56^0$ \\ \hline
0.25 & 1 & 15 & 0.17 & 0.31 & $164^0$ & $53^0$ \\ \hline
0.20 & 1  & 15 & 0.10 & 0.26 & $157^0$ & $50^0$  \\ \hline
0.15 & 1 & 15 & 0.03 &  0.20  & $149^0$ & $48^0$  \\ \hline
\end{tabular}
\end{center}
\end{table}

\begin{multicols}{2}

\noindent As seen from our results, the symmetry of the ground state is
lower than that of the Hamiltonian. The latter has  point group $C_{4d}$
and it has two sets of rotation axis positions: at the centers of the dots
and at the centers of squares formed by the dots. The residual symmetry of
the
ground state is $Z_2$, with reflection in all specific lines of squares.
This
peculiar symmetry violation stems from a highly inhomogeneous magnetic
field
distribution: the screened magnetic field is
logarithmically divergent
near the boundaries of the FM dots. As a result,
the
vortices and the
antivortices are dragged into these regions.

\section{Conclusions}

We considered a periodic array of FM dots on a SC film.  In the
first section, we presented the formalism which is modified for
the case of periodic structures. In this formalism, the problem is
formulated as a variational principle which allows us to calculate
directly the positions of vortices. We applied this formalism to
the case of the square array of the FM dots. We have demonstrated
that the dots can generate the vortices under themselves and
antivortices at interstitial locations. Thus, the symmetry of the
dot array is spontaneously violated by vortices and antivortices.
For different values of the size of the dot and the magnetization,
we calculated the positions of the vortices and the antivortices
and the phase transition curves which separate the regions with
and and without vortex-antivortex pairs. The transition is of the
first order in our approximation. It turns out that, when the
ratio $R/L$ increases, then the minimal value of the ratio $m
\phi_0/\varepsilon_v$ necessary for the spontaneous creation of
vortex-antivortex pairs decreases. We assumed that only one
vortex-antivortex pair per elementary cell of the array is
generated. The problem of many pairs has not yet been solved.

\section{Acknowledgements}
This work is mostly  done in Physics Department at Texas A\&M 
University. We would like to thank V.L.Pokrovsky for the statement of the
problem, and W.M. Saslow for useful discussions.

\end{multicols}


\begin{thebibliography}{10}


\bibitem{e1} J.I. Martin, M. Velez, J. Nogues and I.K. Schuller, Phys.
Rev. Lett. {\bf 79}, 1929 (1997).


\bibitem{e2} D.J. Morgan and J.B. Ketterson Phys.
Rev. Lett. {\bf 80}, 3614 (1998).

 \bibitem{e4} Y. Otani, B. Pannetier, J.P. Nozieres and D.
Givord, J. Magn. Mag. Mat. {\bf 126}, 622 (1993).
\bibitem{e5} O.
Geoffroy, D. Givord, Y. Otani, B. Pannetier and F. Ossart, J. Magn. Mag.
Met. {\bf 121}, 223 (1993).
\bibitem{e6} Y. Nozaki, Y. Otani, K. Runge, H.
Miyajima, B. Pannetier, J.P. Nozieres and G. Fillion, J. Appl. Phys. {\bf
79}, 8571 (1996).
\bibitem{e7} M.J. Van Bael, L. Van Look, K. Temst, M.
Lange, J. Bekaert, U. May, G. Guntherodt, V.V. Moshchalkov and Y.
Bruynseraede, Physica C {\bf 332}, 12 (2000).
\bibitem{e8} A. Terentiev,
D.B. Watkins, L.E. De Long, D.J. Morgan and J.B. Ketterson, Physica C {\bf
332}, 5 (2000).



\bibitem{pok1} I.F. Lyuksyutov and V.L. Pokrovsky, Phys. Rev. Lett. {\bf
81}, 2344 (1998).

\bibitem{pl1} I.F. Lyuksyutov and V.L. Pokrovsky, in {\it Superconducting
Superlattices II: Native and Artificial}, edited by Ivan Bozovic and Davor
Pavuna, SPIE Proceedings Vol. 3480 (SPIE-International Society for Optical
Engineering, Bellingham, WA, 1998), p. 230.

\bibitem{pok3} I.F. Lyuksyutov and D.G. Naugle, Modern Phys. Lett. B {\bf
13}, 491 (1999).

\bibitem{feldman} Feldman D.E., Lyuksyutov I.F., Pokrovsky V.L. and V.M.
Vinokur, Europhys. Lett. {\bf  51}, 110 (2000).

\bibitem{helselth} L.E. Helseth, Phys. Rev. B {\bf 66}, 104508 (2002).


\bibitem{mar} I.K. Marmorkos, A. Matulis and F.M. Peeters, Phys. Rev. B
{\bf 53}, 2677 (1996).

\bibitem{san1} J.E. Santos, E. Frey and F. Schwabl, Phys. Rev. B {\bf 63},
4439 (2001).

\bibitem{amin} M.A. Kayali, Phys. Lett. A {\bf 298}, 432 (2002).

\bibitem{common} S. Erdin, A.M. Kayali, I.F. Lyuksyutov, and V.L.
Pokrovsky, Phys. Rev. B {\bf 66}, 014414 (2002).



\bibitem{ee} M.J. Van Bael, K. Temst, V.V. Moshchalkov and
Y. Bruynseraede, Phys. Rev. B {\bf 59}, 14674 (1999).


\bibitem{met1} V.V. Metlushko, M. Baert, R. Jonckheere, V.V. Moshchalkov
and Y. Bruynseraede, Solid State Comm. {\bf 91}, 331 (1994).
\bibitem{met2} M. Baert, V.V. Metlushko, R. Jonckheere, V.V. Moshchalkov
and Y. Bruynseraede Phys. Rev. Lett. {\bf 74}, 3269 (1995).
\bibitem{met3}
V.V. Moschalkov, M. Baert, V.V. Metlushko, E. Rosseel, M.J. VanBael, K.
Temst, R. Jonckheere and Y. Bruynseraede Phys. Rev. B {\bf 54}, 7385
(1996).
\bibitem{met4} V.V. Metlushko, L.E. DeLong, M. Baert, E. Rosseel,
M.J. VanBael, K. Temst, R. Jonckheere and Y. Bruynseraede, Europhys. Lettl
{\bf 41},333 (1998).

\bibitem{bilayer} S.Erdin, I.F. Lyuksyutov, V.L. Pokrovsky and V.M. 
Vinokur,
Phys. Rev. Lett. {\bf 88}, 017001
(2002).

\bibitem{abrikosov} A.A. Abrikosov, {\it Introduction to the Theory of
Metals}
(North Holland, Amsterdam, 1986).

\end{thebibliography}
\end{document}